\documentclass[12pt]{iopart}
\usepackage{iopams}  
\usepackage{graphicx,bm}

\begin{document}

\title[]{Fulde-Ferrell superfluids in spinless ultracold Fermi gases}
\author{Zhen-Fei Zheng, Guang-Can Guo, Zhen Zheng, and Xu-Bo Zou}
\address{Key Laboratory of Quantum Information, and Synergetic Innovation Center of Quantum Information and Quantum Physics, University of Science and Technology of China, Hefei, Anhui 230026, People's Republic of China}
\ead{zhenzhen@ustc.edu.cn and xbz@ustc.edu.cn}
\vspace{10pt}

\begin{abstract}
The Fulde-Ferrell (FF) superfluid phase, in which fermions form finite-momentum Cooper pairings,
is well studied in spin-singlet superfluids in past decades.
Different from previous works that engineer the FF state in spinful cold atoms,
we show that the FF state can emerge in spinless Fermi gases confined in optical lattice associated with nearest-neighbor interactions.
The mechanism of the spinless FF state relies on the split Fermi surfaces by tuning the chemistry potential,
which naturally gives rise to finite-momentum Cooper pairings.
The phase transition is accompanied by changed Chern numbers,
in which, different from the conventional picture, the band gap does not close.
By beyond-mean-field calculations, we find the finite-momentum pairing is more robust,
yielding the system promising for maintaining the FF state at finite temperature.
Finally we present the possible realization and detection scheme of the spinless FF state.
\end{abstract}

%
\vspace{2pc}

\section{Introduction}

Cold atoms in optical lattices provide an ideal experimental plateau
for quantum simulation of the quantum many-body system.
Compared with conventional solid-state systems,
it possesses remarkable advantages such as the well controllability and tunability of the system parameters
and free of disorder \cite{disorder}.
Furthermore, by utilizing recently developed technique
with laser-assisted tunneling \cite{laser-hop-rev-1,laser-hop-rev-2,laser-hop-exp-1,laser-hop-exp-2}
or periodic-driven external fields \cite{driven-lattice-rev,driven-lattice-scheme-1,driven-lattice-scheme-2,driven-lattice-scheme-3},
cold atoms show promising potential in synthesizing exotic optical lattice models and artificial gauge fields.
It paves the way to quantum simulate various condensed-matter systems,
and search possible unconventional phases that are not ever detected in solid-state systems.
Among them, the Fulde-Ferrell (FF) phase \cite{fflo-1st-ff,fflo-1st-lo} attracts tremendous research interest.

The FF state is an unconventional superfluid state
with spatially oscillating order parameters.
It originates from Cooper pairings with finite center-of-mass momentum,
which is the prominent feature distinguished from Bardeen-Cooper-Schrieffer (BCS) state.
The FF state provides a central concept for understanding exotic phenomena in different physics branches 
\cite{fflo-review,fflo-review-torma}.
It is predicted to emerge in systems with large spin polarization \cite{fflo-h1,fflo-h2,fflo-h3,fflo-h4}.
Due to the stringent conditions on materials, the evidence of the FF state in condensed matters is still pending.
On the other hand, in the past few years,
it opens an alternative way in synthesizing the FF superfluids in cold atoms,
by taking advantages of anisotropic optical lattices \cite{fflo-gauge},
spin-dependent optical lattices \cite{fflo-demler},
spin-orbital couplings (SOC) \cite{fflo-soc1,fflo-soc2,fflo-soc3,fflo-soc4,fflo-soc5,fflo-soc6},
periodic-driven optical lattices \cite{fflo-driven},
multi-orbital interactions \cite{fflo-orbit},
the optical control of Feshbach resonances \cite{fflo-feshbach},
and instantaneous spin imbalance via radio-frequency fields \cite{mueller-fflo}.
The series of investigations in cold atoms reveals that the FF superfluids can originate from
the distortion of Fermi surfaces instead of large spin imbalance \cite{fflo-h2,fflo-h3,fflo-h4,mueller-fflo},
which is expected to facilitate its observation in cold-atom experiments.
As the results, cold atoms exhibit a potential candidate to realize and study the FF superfluids.

So far in searching FF superfluids in cold atoms,
the earlier advances \cite{fflo-gauge,fflo-demler,fflo-soc1,fflo-soc2,fflo-soc3,fflo-soc4,fflo-soc5,fflo-soc6,fflo-driven,fflo-orbit,fflo-feshbach}
bear similarities that they all focus on a spinful system.
In these systems, the contact interaction between opposite pseudo-spin atoms
plays the key role for the superfluid phases.
In cold-atom experiments, the interaction is induced via Feshbach resonance, and conventionally spatial homogeneous.
In order to engineer FF superfluids,
the earlier works design SOC
via current laser techniques to break the homogeneity of the band dispersion,
and thus a distorted Fermi surface is engineered.
However, the idea based on SOC is no longer valid for a spinless system.
For that sake, an interesting question motivates us
whether it is possible to explore FF superfluids in a spinless system.

In this paper, different from previous works that focus on spinful systems,
we show that FF superfluids can emerge in spinless ultracold Fermi gases
trapped in a two-dimensional (2D) optical lattice.
The paper is organized as follows.
In Section \ref{sec-model}, we present the model Hamiltonian and the mean-field framework.
In Section, \ref{sec-res} we show the phase diagram and topological features of the system.
The stability of the emergent FF state against fluctuations
is estimated by the Berezinskii-Kosterlitz-Thouless (BKT) transition temperature \cite{bkt-origin-1,bkt-origin-2,bkt-origin-3},
yielding the system promising for maintaining the FF state at finite temperature.
In Section \ref{sec-diss}, we give a possible scheme to detect the FF state via the pair correlation,
and discuss the experimental realization of the spinless lattice model.
In Section \ref{sec-con}, we summarize the work.

\section{Model Hamiltonian}\label{sec-model}

\begin{figure}[tbp]
\centering
\includegraphics[width=0.85\textwidth]{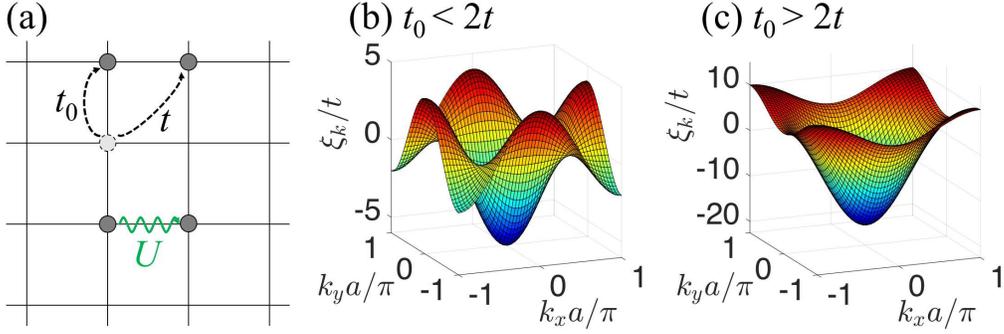}
\caption{(a) Illustration of the lattice model.
(b)-(c) Single-particle band structure in the 1st BZ with (b) $t_0<2t$ and (c) $t_0>2t$.} 
\label{fig-band}
\end{figure}

We start with spinless Fermi gases trapped in a 2D square lattice.
The lattice model is illustrated in Figure \ref{fig-band}(a),
and can be described by the following Hubbard Hamiltonian,
\begin{equation}
H = -t_0\sum_{i,j}\!' c_i^{\dag}c_j -t\sum_{i,j}\!'' c_i^{\dag}c_j - \mu \sum_i \hat{n}_i
- U\sum_{i,j}\!'\hat{n}_i\hat{n}_j~.\label{eq-h}
\end{equation}
Here $c_i^{\dag}$ ($c_i$) are the fermionic creation (annihilation) operators on $i$-th site, respectively,
and $\hat{n}_i\equiv c_i^{\dag}c_i$ are density operators.
The summations $\sum'$ and $\sum''$ range over all nearest-neighbor (NN) and all next-nearest-neighbor (NNN) sites, respectively.
The corresponding tunneling amplitudes are $t_0$ and $t$.
Hereafter we set $t$ as the energy unit.
$\mu$ is the chemistry potential.
$U$ characterizes the attractive interaction strength.
A candidate system described by this model is the fully-spin-polarized Fermi gas with dipole-dipole interactions.
Due to the Pauli exclusion, each site in a spinless lattice system is occupied by a maximum of one fermion.
Therefore the onsite interaction that gives rise to $s$-wave Cooper pairings is prohibited,
while the long-range one is still valid.
In the tight-binding approximation,
our focus here is the NN interaction, which will lead to superfluid order parameters with $p$-wave symmetry \cite{nishida,bwang}.

We firstly investigate the single-particle properties.
The Hamiltonian without interactions in the momentum space is given by
\begin{eqnarray}
H_0(\bm{k}) \equiv \xi_k =& -\mu - 2t_0\cos(k_xa) - 2t_0\cos(k_ya) \nonumber\\
&- 2t\cos(k_xa+k_ya) - 2t\cos(k_xa-k_ya)
~. \label{eq-h0}
\end{eqnarray}
Figure \ref{fig-band}(b)-(c) show the band structures of the single-particle system.
We find that the band hosts five valleys in the center and corners of the first Brillouin zone (BZ) when $t_0<2t$,
by contrast, only one valley is present when $t_0>2t$.
This implies that, by changing $\mu$, the Fermi surface can be split from a single enclosed curve into disjoint lines if $t_0<2t$.
It reveals a possibility for rich Cooper pairing types, inspiring us to search the possible FF state. 
For simplicity, we set $t_0=t/2$ in the calculations of the whole paper.

Then we study the main features of the interacting system at zero temperature.
In order to capture qualitative understanding of the interacting Fermi gas,
we take the mean-field Bogoliubov-de Gennes (BdG) approach to study the superfluid phases.
The order parameter can be introduced by
$-U\hat{n}_i\hat{n}_j = \Delta_i c_i^{\dag} c_{j}^{\dag} + \Delta_i^*c_{j}c_i - |\Delta_i|^2/U$.
Thus the Hamiltonian (\ref{eq-h}) of a $L\times L$ lattice can be diagonalized by employing the Bogoliubov transformation
$c_i = \sum_{\eta}\left(u_i^{\eta}\gamma_{\eta}+v_i^{\eta}\gamma^{\dag}_{\eta}\right)$.
Here $u_{\eta} = \left(u_1^{\eta},\ldots,u_{L\times L}^{\eta}\right)^T$ and $v_{\eta} = \left(v_1^{\eta},\ldots,v_{L\times L}^{\eta}\right)^T$
satisfy the following BdG equations,
\begin{equation}
\left( \begin{array}{cc}
\hat{h} & \hat{\Delta} \\
\hat{\Delta}^\dag & -\hat{h}
\end{array}\right)
\left(\begin{array}{c}
u_{\eta} \\
v_{\eta}^*
\end{array}\right) =
E_{\eta} \left(\begin{array}{c}
u_{\eta} \\
v_{\eta}^*
\end{array}\right) ~, \label{eq-h-bdg-real}
\end{equation}
$E_{\eta}$ is the excitation energy for the $\eta$-th quasiparticle state,
$\hat{h}_{ij} = -\mu\delta_{ij} -t_0\delta_{\bm{i},\bm{j}\pm\hat{\bm{e}}_{x/y}}-t\delta_{\bm{i},\bm{j}\pm\hat{\bm{e}_{x}}\pm\hat{\bm{e}}_{y}}$,
and $\hat{\Delta}_{ij} = \Delta_i\delta_{\bm{i},\bm{j}-\hat{\bm{e}}_{x/y}} - \Delta_{i-1}\delta_{\bm{i},\bm{j}+\hat{\bm{e}}_{x/y}}$.
Here $\hat{\bm{e}}_{x/y}$ is denoted as the lattice vector basis along the $x$/$y$ direction.

We numerically solve Eq. (\ref{eq-h-bdg-real}) to self-consistently determine $\Delta_i$ at a fixed $\mu$
with a periodic boundary condition.
The ground state is determined by comparing results obtained by randomly choosing initialized configurations.
When $\{\Delta_i\}$ is a nonzero constant,
the system is in a BCS state.
When $\{\Delta_i\}$ hosts a spatially periodic structure, \textit{i.e.} the FF-type pairing, the system is in an FF state \cite{fflo-1st-ff}.
When $\{\Delta_i\}$ vanishes, the system is a trivial normal gas (NG) state.

\begin{figure}[tbp]
\centering
\includegraphics[width=0.9\textwidth]{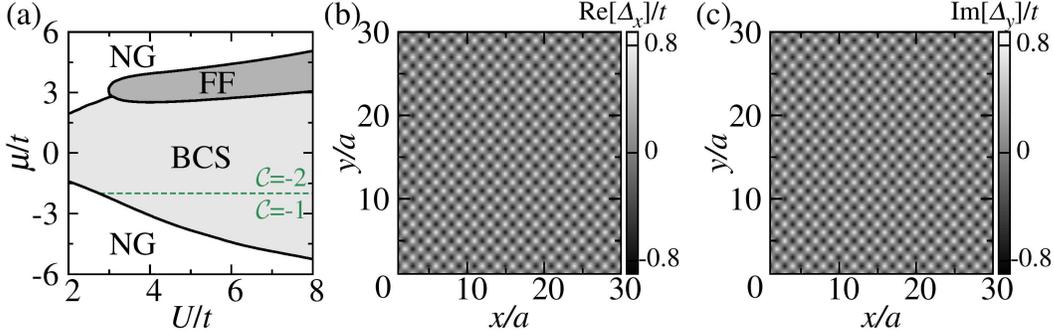}
\caption{(a) Phase diagram in the $U$-$\mu$ plane at zero temperature.
The green dashed line marks the change of the Chern number in the BCS phase region.
(b)-(c) Spatial distribution of the order parameters in the FF superfluids.
The gray level visualizes the value of the order parameters 
$\mathrm{Re}[\Delta_x]$ and $\mathrm{Im}[\Delta_y]$, respectively.
Results in (b)-(c) are obtained for $U = 6.0t$ and $\mu=3.5t$ on a 30$\times$30 lattice. $a$ is the lattice constant.}
\label{fig-phase}
\end{figure}

\section{Results}\label{sec-res}

\subsection{Phase diagram}\label{sec-phase}

In Figure \ref{fig-phase}(a), 
we plot the phase diagram at zero temperature with respect to two experimentally tunable parameters: 
the chemistry potential $\mu$ and the interaction strength $U$.
We find that the FF state appears in the high filling regime (characterized by large $\mu$)
when $U$ exceeds a critical value in BCS--Bose-Einstein-condensation(BEC) crossover.
This is different from the picture of the spin polarized system \cite{fflo-torma},
in which the FF state exists in both high and low filling regime.
To capture the feature of the FF superfluids, we plot the spatial dependence of the order parameters in Figure \ref{fig-phase}(b)-(c).
In the 2D system, the order parameters can be separated into two parts:
$\{\Delta_{x}\}$ and $\{\Delta_{y}\}$, which characterize the pairing between two adjacent sites along $x$ and $y$ directions, respectively.
Their magnitudes are identical due to the homogeneity of $x$ and $y$ directions,
but their phases host a relative $\pi/2$ difference \cite{p-wave-n1,p-wave-n2,p-wave-n3}.
For simplicity without loss of generality, we assume $\mathrm{Im}[\Delta_x]=\mathrm{Re}[\Delta_y]=0$
where $\mathrm{Im}[z]$ and $\mathrm{Re}[z]$ is the imaginary and real part of a complex number $z$, respectively.
In Figure \ref{fig-phase}(b)-(c), we see that $\{\Delta_{x}\}$ and $\{\Delta_{y}\}$ acquire a spontaneous spatial-modulated phase and
vary individually like $\mathrm{Re}[\Delta_{x,\bm{j}}]=\Delta e^{\mathrm{i}\bm{Q}\cdot \bar{\bm{r}}_{x,\bm{j}}}$
and $\mathrm{Im}[\Delta_{y,\bm{j}}]=\Delta e^{\mathrm{i}\bm{Q}\cdot \bar{\bm{r}}_{y,\bm{j}}}$,
with a periodic $\bm{Q}=(\pi/a,\pi/a)$.
Here $a$ is the lattice constant.
$\bar{\bm{r}}_{x/y,\bm{j}}=(\bm{r}_{\bm{j}}+\bm{r}_{\bm{j}+\hat{\bm{e}}_{x/y}})/2$
is the center-of-mass coordinate of the Cooper pairings.
It implies $\{\Delta_{x}\}$ and $\{\Delta_{y}\}$ host a checkerboard structure in the real space.

\begin{figure}[tbp]
\centering
\includegraphics[width=0.8\textwidth]{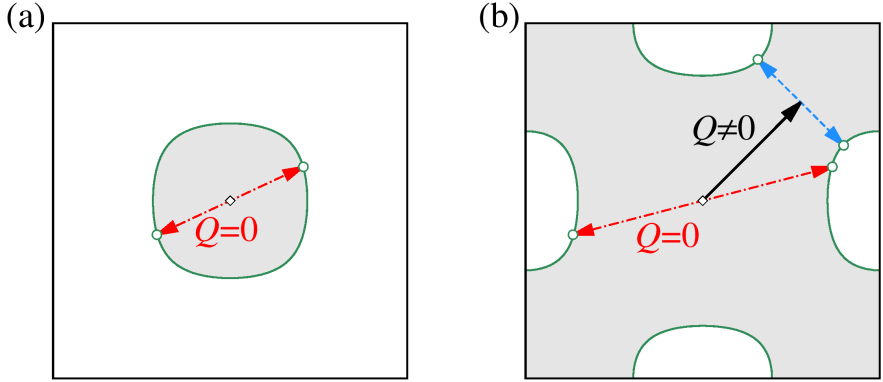}
\caption{Illustration of the Fermi surfaces in the 1st BZ at (a) $\mu=-2.0t$ and (b) $\mu=3.0t$.
Fermions occupy the grey regions enclosed by the Fermi surfaces.
In the low filling regime (a), the system hosts a single Fermi surface (green solid boundaries of the gray region),
and the BCS-type pairing (red dash-dotted arrows) is dominant.
In the high filling regime (b), the Fermi surfaces are split into four disjoint lines in the 1st BZ.
It leads to a competition between the BCS-type pairing,
and the FF-type one (blue dashed arrows) with a finite momentum $\bm{Q}$ (black solid arrows).} 
\label{fig-fs}
\end{figure}

To understand the physical mechanism of the emergent FF superfluids,
in Figure \ref{fig-fs}, we illustrate the Fermi surfaces of the single-particle Hamiltonian (\ref{eq-h0}) in the 1st BZ.
The NNN tunneling $t$ can induce a deformation to the Fermi surfaces, which brings out the possibility to search the finite-momentum pairing.
In the low filling regime, the system hosts a single Fermi surface,
thus the zero-momentum pairing is dominant.
By contrast, in the high filling regime, the Fermi surfaces are split into four disjoint sectors.
It will lead to a competition between two types of possible Cooper pairings ---
the zero/finite-momentum pairings.
Due to the homogeneity along the $x$ and $y$ directions, the FF state acquires a finite momentum $\bm{Q}$ along ($\pm1,\pm1$) directions.
In Figure \ref{fig-phase}(a) in the high filling regime, we see that the increase of $\mu$ drives a transition from the BCS to FF states.
It yields in the large-$\mu$ regime, the finite-momentum pairing dominates over the zero-momentum one.

\subsection{Topological phase transition}\label{sec-topo}

The phase diagram shown in Figure \ref{fig-phase}(a) 
is accompanied by the topological transition.
It has been well studied that 
the chiral $p$-wave superfluids can exhibit features of a Chern insulator,
and harbor the topological edge states protected by the particle-hole symmetry \cite{topo-cri,topo-cri-1,topo-cri-2,topo-cri-3}.
The topological phase transition can be characterized by the Chern number $\mathcal{C}$.
It is defined by \cite{chern-num}
\begin{equation}
\mathcal{C} = \frac{1}{2\pi} \int\mathrm{d}\bm{k} \, \mathrm{Tr} F_{xy}(\bm{k}) ~,
\end{equation}
where the gauge field $F_{xy}(\bm{k})= \partial_{k_x} A_y(\bm{k}) - \partial_{k_y} A_x(\bm{k})$,
the Berry connection $A_{\mu}(\bm{k})=-\mathrm{i}\langle \alpha_{\bm{k}}|\partial_\mu | \alpha_{\bm{k}}\rangle$,
and $|\alpha_{\bm{k}}\rangle=(|\alpha_{1\bm{k}}\rangle,\cdots,|\alpha_{\eta\bm{k}}\rangle,\cdots)^T$
with $|\alpha_{\eta\bm{k}}\rangle$ as the base of the $\eta$-th occupied band.
$|\alpha_{\eta\bm{k}}\rangle$ can be obtained by the BdG Hamiltonian,
which, in the base $\Psi=(c_{\bm{Q}/2+\bm{k}},c_{\bm{Q}/2-\bm{k}}^\dag)^T$, is expressed as
\begin{equation}
H_\mathrm{BdG}(\bm{k})=
\left(\begin{array}{cc}
\xi_{\bm{Q}/2+\bm{k}} & \Delta_{\bm{k}}\mathrm{e}^{\mathrm{i}\varphi/2} \\
\Delta^\dag_{\bm{k}}\mathrm{e}^{-\mathrm{i}\varphi/2} & -\xi_{\bm{Q}/2-\bm{k}}
\end{array}\right) ~,
\label{eq-h-bdg}
\end{equation}
where $\Delta_{\bm{k}} = \mathrm{i} 2\Delta\sin(k_xa) + 2\Delta\sin(k_ya)$.
For simplicity, we have denoted $\varphi=|\bm{Q}_{x/y}|a=l\pi$ with $l=0(1)$ for the BCS(FF) state, respectively.

\begin{figure}[tbp]
\centering
\includegraphics[width=0.85\textwidth]{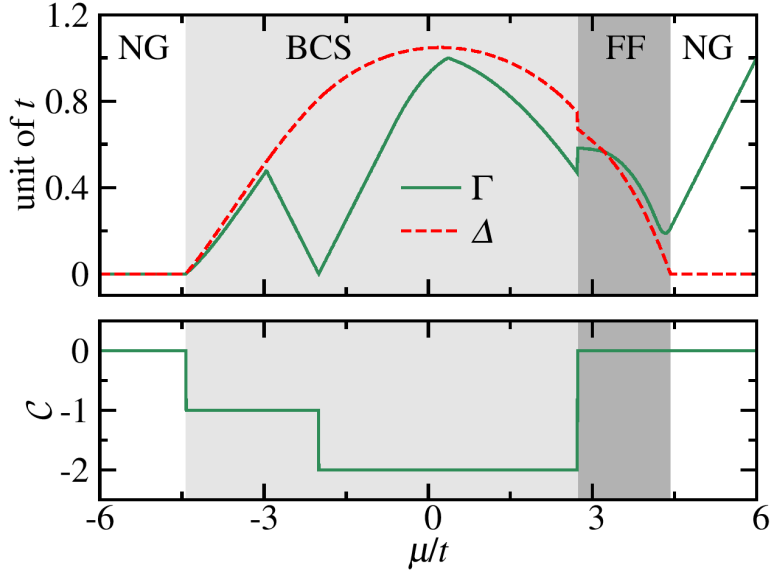}
\caption{The band gap $\Gamma$, the order parameter $\Delta$, and the Chern number $\mathcal{C}$ as a function of $n$.
Results are obtained by setting $U=6.0t$.} 
\label{fig-topo}
\end{figure}

In Figure \ref{fig-topo} we plot the order parameter $\Delta$, the band gap $\Gamma$,
and the Chern number $\mathcal{C}$ with respect to $\mu$.
Here $\Gamma$ is defined by $\Gamma= \mathrm{min}|E_{\bm{k}}|$ 
($E_{\bm{k}}$ is the eigenvalue of the Hamiltonian (\ref{eq-h-bdg})).
It describes the gap between particle and hole bands.
In the low filling regime (small $\mu$), although the gap $\Gamma$ closes and reopens when increasing $\mu$,
the system is still the topological $p_x$+i$p_y$ wave superfluids with nonzero $\mathcal{C}$.
It is easy to be demonstrated that $\mathcal{C}$ changes at $\mu=-2t$, 
since $\Gamma$ can vanish at the corners of the 1st BZ.
Therefore, the boundary of the two BCS phases is independent from $U$, which has been shown in Figure \ref{fig-phase}(a).
In the high filling regime (large $\mu$), the system processes a transition from the BCS to FF phase, 
which spontaneously breaks the translational invariance of the superfluid order $\Delta$.
The phase transition also undergoes a change of $\mathcal{C}$, however, the gap $\Gamma$ is still open during the transition.
This is because the transition is of first order, which is revealed by the discontinuous behavior of $\Delta$ with respect to $\mu$.
The evolution from the topologically nontrivial BCS ($\mathcal{C}\neq0$) 
to topologically trivial FF phase ($\mathcal{C}=0$) is thus not an adiabatic continuum deformation.

\subsection{Stability of FF superfluids}\label{sec-stability}

\begin{figure}[tbp]
\centering
\includegraphics[width=0.9\textwidth]{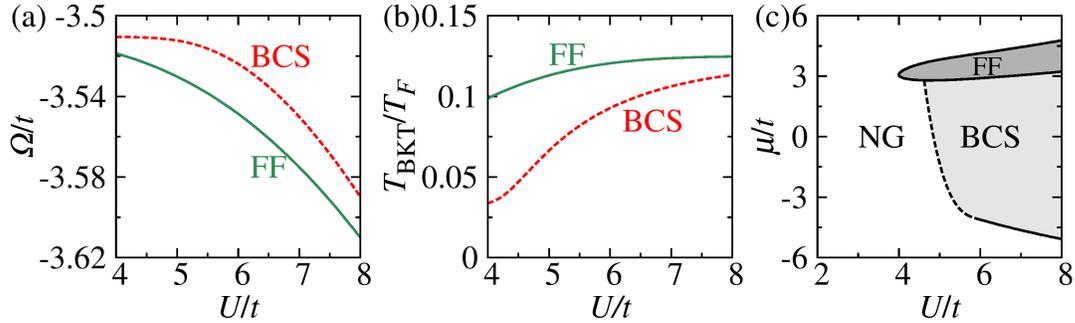}
\caption{(a) Thermodynamic potential $\Omega$ of two pairing states 
as a function of $U$ for $\mu=3.0t$.
The critical points of the phase transition are marked by black empty diamonds.
(b) BKT transition temperature of two pairing states  as a function of $U$ at for $\mu=3.0t$.
$T_F=\pi n/m$ is the Fermi temperature (see \ref{app-fluctuation}).
(c) Phase diagram in the $U$-$\mu$ plane at the temperature $T=0.05T_F$.
The dash line is self-consistently obtained by setting $T_\mathrm{BKT}=0.05T_F$,
characterizing the superfluid phase transition caused by the phase fluctuations.}
\label{fig-bkt}
\end{figure}

At zero temperature, in Figure \ref{fig-bkt}(a), we calculate 
the difference of the thermodynamic potential $\Omega$ between the FF and possible BCS states (see \ref{app-mean-field}).
We can see that the FF state hosts lower energy than the possible BCS state.
It reveals the dominance of the finite-momentum pairing over the zero-momentum one in the high filling regime,
thus the FF state is the ground state.

At finite temperature,
the long-range superfluid order in a 2D system is destroyed by its phase fluctuations.
Instead, states with quasi-long-range order, which are characterized by the vortex-antivortex pairs \cite{bkt1,bkt2,bkt3,bkt4},
drive a BKT-type phase transition.
The superfluids are formed below the critical temperature, which is known as the BKT transition temperature $T_\mathrm{BKT}$.
When the temperature exceeds $T_{\mathrm{BKT}}$, the ground state of the system turns to the pseudo-gap phase,
in which the superfluid components are destroyed even though the pairings $\{\Delta\}$ do not vanish.
The stability of the FF superfluids at finite temperature can therefore be estimated by $T_\mathrm{BKT}$.

In order to study the phase fluctuations,
we impose a phase $\theta$ in the superfluid order parameters $\Delta=\Delta e^{\mathrm{i}\theta}$.
After making the standard Hubbard-Stratonovich transformation and integrating out the fermion fields $\{c,c^\dag\}$ (see \ref{app-mean-field}),
the effective action can be expressed as $S_\mathrm{eff}= S_0 + S_{\mathrm{fl}}$.
$S_0$ describes the mean-field action independent from $\theta$.
$S_{\mathrm{fl}}$ characterizes the $\theta$-dependent action originated from the phase fluctuations.
Its form is written as
$ S_{\mathrm{fl}}=\frac{1}{2}\int \mathrm{d}\bm{r}\mathrm{d}\tau
\sum_{\mu,\nu=x,y} \big[ ( J_{\mu\nu}\partial_\mu\theta\partial_\nu\theta
+\mathrm{i}J_{\tau \nu}\partial_{\tau}\theta\partial_\nu\theta )
+ P(\partial_\tau \theta)^2 -\mathrm{i}A\partial_\tau \theta \big]$.
The detailed derivations of $J_{\mu\nu}$, $P$, and $A$ are presented in \ref{app-fluctuation}.
The BKT transition temperature is then determined by \cite{bkt4}
\begin{equation}
T_\mathrm{BKT}=\frac{\pi}{2}\sqrt{J_{xx}J_{yy}}~. \label{eq-t-bkt}
\end{equation}

In Figure \ref{fig-bkt}(b) we plot $T_\mathrm{BKT}$ of the FF and possible BCS states by changing the interaction strength $U$.
We see that the FF superfluids still exist and remain robust against the phase fluctuations below $T_{\mathrm{BKT}}$,
yielding the system promising for maintaining the FF superfluids at finite temperature.
In the BCS regime, $T_{\mathrm{BKT}}$ increases monotonically to the interaction strength $U$.
However, in the BEC regime,
$T_{\mathrm{BKT}}$ approaches a constant independent from $U$.
This is because the system behaves like a condensation of tightly-bound bosonic dimers due to the strong attractive interaction \cite{bkt1}.
Since $T_\mathrm{BKT}$ of the FF state is higher than the possible BCS state,
it implies the finite-momentum pairing can enhance the superfluids robust against the fluctuations.
We gives the finite-temperature phase diagram in Figure \ref{fig-bkt}(c).
Compared with the zero-temperature one (see Figure \ref{fig-phase}(a)),
it displays the BCS phase region shrinks obviously, while the FF phase region changes slightly.

\section{Discussions}\label{sec-diss}

\subsection{Pair Correlation}\label{sec-pair-corr}

The signature of the FF superfluids can be detected by the pair correlations \cite{correlation}.
At the critical transition point, the pair correlations will exhibit a discontinuous behavior by tuning $\mu$.
In Figure \ref{fig-topo}(a), 
we have known that the order parameter $\Delta$ hosts a discontinuous evolution
during the transition from the BCS to FF states.
This discontinuous behavior will influence on the pair correlations \cite{correlation}.

For a spinless Fermi gas,
the pair wave function between the $i$-th and $j$-th sites is expressed as
\begin{equation}
\Delta(\bm{r}_i, \bm{r}_j) = \langle c(\bm{r}_i) c(\bm{r}_j) \rangle ~.
\end{equation}
By making the following transformation,
\begin{equation}
\bm{r}_c=(\bm{r}_i + \bm{r}_j)/2 ~,\qquad
\delta \bm{r}=\bm{r}_i - \bm{r}_j ~,
\end{equation}
the pair wave function $\Delta(\bm{r}_i, \bm{r}_j)$ can be rewritten as $\Delta(\bm{r}_c, \delta\bm{r})$ in the center-of-mass frame.
The mean pair correlation function can thus be obtained by \cite{correlation,pair-njp}
\begin{equation}
P=\frac{1}{N_L}\sum_{\bm{r}_c,\delta\bm{r}}\big|\Delta(\bm{r}_c, \delta\bm{r})\big|^2 ~,
\end{equation}
where $N_L=L\times L$ is the total number of the 2D lattice sites.

\begin{figure}[tbp]
\centering
\includegraphics[width=0.85\textwidth]{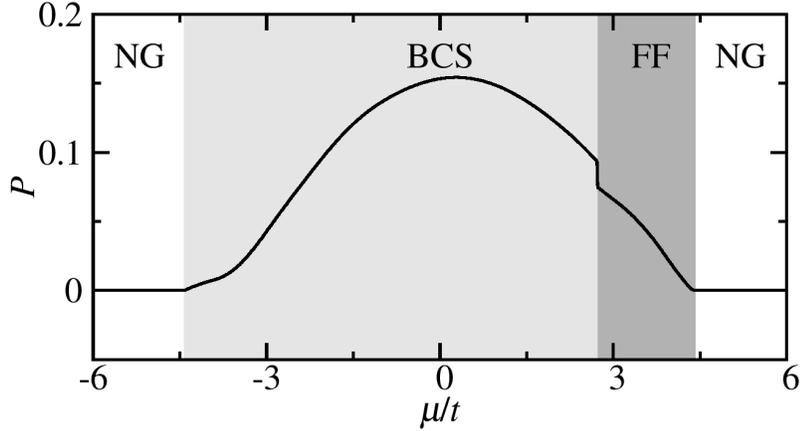}
\caption{Mean pair correlation function $P$ with respect to $\mu$.
Numeric results are obtain on a $30\times30$ lattice with $U=6.0t$.}
\label{fig-crl}
\end{figure}

In Figure \ref{fig-crl}, we plot $P$ with respective to $\mu$ and see that
$P$ behaves a sudden jump at the transition critical points.
It implies us a possible way to detect the phase transition from the BCS to FF states
via current experimental techniques \cite{fflo-h4}.

\subsection{Experimental realization}\label{sec-exp}

The $p$-wave superfluids in a 2D optical lattice can be readily designed by various proposals in cold atoms.
The NN and NNN tunnelings can be constructed by the laser-assisted tunneling protocol \cite{laser-hop-theo}.
Here we give two possible schemes.
The first scheme is that we can introduce a magnetic gradient field to generate adjacent-site detuning $\delta_{\bm{j}}=j_x\delta_x+j_y\delta_y$.
Here $\bm{j}=(j_x,j_y)$ denotes the site index in the 2D lattice.
Then we can implement Raman transitions with detuning $\delta_{x/y}$ to generate the NN tunneling $t_0$,
and ones with detuning $\delta_x\pm\delta_y$ to generate the NNN tunneling $t$.
The tunneling amplitudes can be changed separately.
An alternative scheme takes advantages of the magnetic gradient field with checkerboard structure $\delta_{\bm{j}}=(-1)^{j_x+j_y}\delta$.
The NN tunneling $t_0$ is prohibited since the adjacent-site detuning is $2\delta$,
making the NNN tunneling $t$ is prominent because of no detuning between two NNN sites.
Then $t_0$ is reconstructed by a Raman transition with detuning $2\delta$.

The spin-triplet interaction in the spinless system can be introduced directly via $p$-wave Feshbach resonances \cite{p-wave-0,p-wave-2018}.
Several works provide alternative ways for synthesizing $p$-wave superfluids
by implementing artificially induced molecules \cite{p-wave-1}, higher orbital atoms \cite{p-wave-2},
or Bose-Fermi mixture \cite{p-wave-3}.

\subsection{Large-$U$ limit}\label{sec-largeU}

In this paper, we focus on the BCS-BEC crossover with $\Delta\sim t$
because the mean-field method can give a good picture within the interaction strength we choose \cite{crossover}.
However, when $U$ is extremely large with $\Delta\gg t$, fermions are tightly bounded into bosonic molecules
\cite{composite-boson}.
The system will exhibit a hard-core bosonic gas with tiny tunneling and strong long-range attractive interaction,
yielding a Mott insulator \cite{mott-sf} whose dispersion is nearly a flat band.
At this time, it should be noted that the mean-field method is no longer appropriate to describe the system.

\section{Conclusions}\label{sec-con}

In summary, we investigate the emergent FF superfluids in a spinless Fermi gas.
The novelty of our work is highlighted as follows:
(i) The FF state is supported by the split Fermi surfaces other than the spin imbalance.
Different from the spinful case, 
the Cooper pairing momentum does not depend on external polarized fields.
(ii) The order parameters of the FF state stems from the $p$-wave symmetric pairing,
and forms a checkerboard spatial structure.
(iii) The topological phase transition between the BCS and FF states is of first order,
thus can occur even without gap closing.
(iv) By employing the beyond-mean-field analysis,
we find the finite-momentum pairing is more robust against phase fluctuations than the zero-momentum one.
(v) The lattice model and the associated FF state 
are readily realized and detected via current experimental techniques in cold atoms.
These features above distinguish our work from the conventional pictures for the mechanics and properties of the FF state,
showing the lattice model as a promising candidate system for evidencing and investigating the FF state in cold atoms.

\section{Acknowledgements}

This work is supported by National Natural Science Foundation of China (Grants No. 11674305 and No. 11474271),
Young Scientists Fund of the National Natural Science Foundation of China (Grant No. 11704367),
and National Postdoctoral Program for Innovative Talents of China (Grant No. BX201600147).

\appendix

\section{Mean-Field Approach}\label{app-mean-field}

The Hamiltonian of the 2D model in Section \ref{sec-model} can be expressed as
\begin{equation}
H(\bm{r}) = c^\dag(\bm{r}) \Big[ -\frac{\hbar^2}{2m}\nabla^2 + V_\mathrm{trap}(\bm{r}) \Big] c(\bm{r}) 
- U c^\dag(\bm{r})c^\dag(\bm{r}+\delta\bm{r})c(\bm{r}+\delta\bm{r})c(\bm{r}) \,.
\end{equation}
Here $V_\mathrm{trap}(\bm{r})=V_0\cos^2(x/a)+V_0\cos^2(y/a)$ with the lattice constant $a$ is the lattice trap potential.
Hereafter we set $\hbar=1$ and the Boltzmann constant $k_B=1$.
The partition function is given by
\begin{equation}
Z=\int D[c,c^\dag]e^{-S_{\rm eff}[c,c^\dag ]}
\end{equation}
with the effective action formulated as \cite{stoof-book}
\begin{equation}
S_{\rm eff}[c,c^\dag] = \int \mathrm{d}\tau\mathrm{d} \bm{r} \,
\big[ c^\dag(\bm{r},\tau) \partial_\tau c(\bm{r},\tau) - H(\bm{r}) \big] ~.
\end{equation}
Here $\tau$ is the imaginary time.
We employ the standard Hubbard-Stratonovich transformation with the following pairing fields
\begin{eqnarray}
\Delta_x(\bm{r}) = U\langle c(\bm r)c(\bm r + \delta x)\rangle = 
\Delta e^{\mathrm{i}\bm{Q}\cdot\bar{\bm{r}}_x} ~,\\
\Delta_y(\bm{r}) = U\langle c(\bm r)c(\bm r + \delta y)\rangle =
\Delta e^{\mathrm{i}\bm{Q}\cdot\bar{\bm{r}}_y}
~,
\end{eqnarray}
where
\begin{equation}
\bm{Q}=(\pi/a,\pi/a)~,\quad \bar{\bm{r}}_\nu= \bm{r}+\delta \nu/2 ~,~(\nu=x/y)~.
\end{equation}
Integrating out the fermion fields $\{c,c^\dag\}$,
we obtain the effective action expressed as
\begin{equation}
S_{\rm eff} [\Delta,\Delta^\dag] = \int {\rm d}\tau {\rm d} {\bm r} \Big[ \epsilon_0({\bm r}) 
 -\frac{1}{2} \mathrm{Tr}\ln G^{-1}(\bm{r},\tau) \Big] ~. \label{eq-sm-s}
\end{equation}
Here $G^{-1}(\bm{r},\tau) = -\partial_{\tau} - H_\mathrm{BdG}(\bm r)$.
$\epsilon_0 = |\Delta_x(\bm{r})|^2/U + |\Delta_y(\bm{r})|^2/U -\mu/2$.
$G^{-1}(\bm{r},\tau)$ is the inverse Green's function.

We make the Fourier transformation from the $(\bm{r},\tau)$ space to the $(\bm{k},\mathrm{i}\omega_n)$ space.
Here $\omega=(2n+1)\pi /\beta$ ($n\in\mathbb{Z}$) is the fermionic Matsubara frequency,
and $\beta\equiv 1/T$ with the temperature $T$.
By choosing the base $\Psi_{\bm{k}}=(c_{\bm{Q}/2+\bm{k}},c_{\bm{Q}/2-\bm{k}}^\dag)^T$,
the BdG Hamiltonian $H_\mathrm{BdG}$ in the tight binding approximation with  is expressed as
\begin{equation}
H_\mathrm{BdG}(\bm{k})=
\left(\begin{array}{cc}
\xi_{\bm{Q}/2+\bm{k}} & \Delta_{\bm{k}}\mathrm{e}^{\mathrm{i}\varphi/2} \\
\Delta^\dag_{\bm{k}}\mathrm{e}^{-\mathrm{i}\varphi/2} & -\xi_{\bm{Q}/2-\bm{k}}
\end{array}\right) ~.\label{eq-sm-hamit}
\end{equation}
The thermodynamical potential is written as \cite{mean-field-energy}
\begin{equation}
\Omega = \epsilon_0 -\frac{1}{2\beta}\sum_{\bm k, \mathrm{i}\omega_n,\alpha} 
\ln\big[-\beta(\mathrm{i}\omega_n
- E_{\bm k}^{\alpha})\big]
= \epsilon_0 -\frac{1}{2\beta}\sum_{\bm{k},\alpha} \ln\big(1+e^{-\beta E_{\bm k}^{\alpha}}\big) \,.
\end{equation}
where $E_{\bm k}^{\alpha}$ is the $\alpha$-th eigenvalue of the Hamiltonian (\ref{eq-sm-hamit}),
and $\epsilon_0 = \sum_k\big(|\Delta_x|^2/U + |\Delta_y|^2/U + \xi_k/2\big)$.
The filling factor $n$ can be obtained by
\begin{equation}
n=-\frac{\partial \Omega}{\partial \mu} ~.
\end{equation}

\section{Phase Fluctuation}\label{app-fluctuation}

In order to calculate the phase fluctuation in the 2D system,
we impose a variable phase in the order parameter $\Delta \rightarrow \Delta\mathrm{e}^{i\theta}$ in the Hamiltonian (\ref{eq-sm-hamit}).
By making the following unitary transformation \cite{bkt-torma}
\begin{equation}
U = \left(\begin{array}{cc}
e^{\mathrm{i}\theta/2} & 0 \\
0 & e^{-\mathrm{i} \theta/2}
\end{array}\right) ~,
\end{equation}
the inverse Green's function $\mathcal{G}^{-1}(\bm{r},\tau,\theta)$ in the new representation can be divided into two items,
\begin{equation}
\mathcal{G}^{-1}(\bm{r},\tau,\theta) =
U^\dag \big[ -\partial_{\tau} - H_\mathrm{BdG}(\bm r,\theta) \big] U 
= G^{-1}(\bm{r},\tau) - \Sigma(\bm{r},\tau,\theta) ~.
\end{equation}
The first item $G$ is the original $\theta$-independent Green's function.
Its form in the momentum space is given by 
\begin{equation}
G^{-1}(\bm{k},\mathrm{i}\omega_n)=\mathrm{i}\omega_n - H_\mathrm{BdG}(\bm{k}) ~.
\end{equation}
The second item $\Sigma$ is the $\theta$-dependent self energy expressed as \cite{pseudogap-review}
\begin{equation}
\Sigma = \big[\mathrm{i}\partial_\tau \theta/2+(\nabla\theta)^2/8m\big]\sigma_z
-(\mathrm{i}\nabla^2\theta/4m + \mathrm{i}
\nabla\theta\cdot\nabla/2m)\mathbb{I} ~.
\end{equation}
Here $\sigma_i~(i=x,y,z)$ are Pauli matrices, and $\mathbb{I}$ is the 2$\times$2 identical matrix. 
The effective action (\ref{eq-sm-s}) now becomes
\begin{equation}
S_{\rm eff} =  \int {\rm d}\tau {\rm d} {\bm r} \Big[ \epsilon_0({\bm r}) 
-\frac{1}{2} \mathrm{Tr}\ln \mathcal{G}^{-1} \Big]
\equiv S_0 + S_\mathrm{fl}
\end{equation}
with
\begin{eqnarray}
& S_0 = \int {\rm d}\tau {\rm d} {\bm r} \Big[ \epsilon_0({\bm r})
-\frac{1}{2} \mathrm{Tr}\ln G^{-1} \Big] ~,\\
& S_\mathrm{fl} = -\frac{1}{2}\int {\rm d}\tau {\rm d} {\bm r} \,\mathrm{Tr}\ln(1-G\Sigma) 
~.
\end{eqnarray}
We expand $S_\mathrm{fl}$ to the second order and obtain
\begin{eqnarray}
S_\mathrm{fl} \approx \frac{1}{2} \int {\rm d}\tau {\rm d} {\bm r}
\Big[\mathrm{Tr}(G\Sigma) + \frac{1}{2} \mathrm{Tr}(G\Sigma G\Sigma)\Big] \nonumber\\
= \frac{1}{2}\int \mathrm{d}\bm{r}\mathrm{d}\tau \Big[
\sum_{\mu,\nu=x,y} ( J_{\mu\nu}\partial_\mu\theta\partial_\nu\theta
+\mathrm{i}J_{\tau \nu}\partial_{\tau}\theta\partial_\nu\theta )
+ P(\partial_\tau \theta)^2 -\mathrm{i}A\partial_\tau \theta \Big] \,,
\end{eqnarray}
where
\begin{eqnarray}
& J_{\nu\nu} = \frac{n}{4m} + \frac{\beta}{8}\sum_{\bm{k},\alpha}
\frac{k_\nu^2}{m^2} f(E^{\alpha}_{\bm{k}})\big[f(E^{\alpha}_{\bm{k}})-1\big]
\,,\quad J_{xy} = 0 \,,\label{eq-sm-jxx}\\
& J_{\tau \nu} = \frac{1}{(2\pi)^2}\frac{\beta}{4}\sum_{{\bm k},\mathrm{i}\omega_n} 
\frac{k_\nu}{m}{\rm Tr} \big[ G(\bm k, \mathrm{i}\omega_n)\, 
G(\bm k, \mathrm{i}\omega_n) \, \sigma_z \big] \,,\\
& P = - \frac{1}{(2\pi)^2}\frac{\beta}{8}\sum_{{\bm k},\mathrm{i}\omega_n} {\rm Tr} 
\big[ G(\bm k, \mathrm{i}\omega_n)\,\sigma_z \,
G(\bm k, \mathrm{i}\omega_n)\,\sigma_z \big] \,,\\
& A = n \,,
\end{eqnarray}
and $f(E)=\frac{1}{e^{\beta E}+1}$ is the Fermi-Dirac distribution.
The BKT transition temperature $T_\mathrm{BKT}$ 
is determined by self-consistently solving the following equations
\begin{equation}
T_\mathrm{BKT} = \frac{\pi}{2}\sqrt{J_{xx}J_{yy}} \,,~
\frac{\partial }{\partial \Delta} \Omega(\Delta,\mu,T_\mathrm{BKT}) 
= \frac{\partial }{\partial \mu} \Omega(\Delta,\mu,T_\mathrm{BKT})=0
\,.
\end{equation}
We note that, in the BEC regime, $\Delta\gg t$ leads to $f(E^{\alpha}_{\bm{k}})\big[f(E^{\alpha}_{\bm{k}})-1\big]\sim 0$ in Eq. (\ref{eq-sm-jxx}).
Therefore $J_{xx} = J_{yy} \approx \frac{n}{4m} $ and hence
\begin{equation}
T_\mathrm{BKT}\approx\frac{\pi}{8m} n ~. \label{eq-sm-t-bkt}
\end{equation}
This is different from the three-dimensional (3D) case,
in which the superfluid transition temperature $T_\mathrm{c}\sim n^{2/3}$ \cite{t-matrix}.

It should be noted that in the lattice calculations at nonzero temperature, since we have assume $\hbar=k_B=m=1$,
it is no longer appropriate to use the tunneling amplitude $t$ as the temperature unit.
Instead, we use the Fermi temperature $T_F$ as the temperature unit.
In particular, for the calculation of Figure \ref{fig-bkt},
we have set $t=0.08E_R$ with the lattice recoil energy $E_R=\hbar^2k_L^2/2m$ ($k_L\equiv\pi /a$).

$T_F$ can be defined as following.
In the 2D system, the Fermi wave vector $k_F$ is obtained by
\begin{equation}
n=\frac{N}{S}=\frac{2\times \pi k_F^2}{(2\pi)^2} ~. \label{eq-sm-kf}
\end{equation}
The Fermi temperature $T_F$ is given by
\begin{equation}
T_F=\frac{k_F^2}{2m} ~. \label{eq-sm-tf}
\end{equation}
Combining Eqs. (\ref{eq-sm-kf}) and (\ref{eq-sm-tf}), we obtain the final expression of the Fermi temperature:
\begin{equation}
T_F=\pi n/m ~. \label{eq-sm-tf2}
\end{equation}
Inserting Eq. (\ref{eq-sm-tf2}) into Eq. (\ref{eq-sm-t-bkt}), 
we can obtain the relation between $T_\mathrm{BKT}$ in the BEC regime and $T_F$,
\begin{equation}
T_\mathrm{BKT}\approx T_F/8 ~.
\end{equation}
By contrast, the ratio $T_\mathrm{c}/T_F\approx 0.218$ for the 3D case \cite{t-matrix}.
The practical temperature in cold-atom experiments is typically of order $10^{-2}T_F$ \cite{exp-rev}.
Therefore, it is expected that the FF superfluids, or the nonzero pairing momentum state, can exist in real experiments.

\section*{References}
\bibliography{ref}

\end{document}